\documentclass[aps, prc, reprint]{revtex4-2} 

\usepackage[T1]{fontenc} 
\usepackage[english]{babel} 
\usepackage{amsmath,amsfonts,amsthm} 
\usepackage{gensymb}
\usepackage{textcomp}
\usepackage{verbatim}

\usepackage{graphicx}

\usepackage[]{hyperref}  

\begin{document}


\title{Influence of the recoil-order and radiative correction on the beta decay correlation coefficients in mirror decays}
\author{S.~Vanlangendonck}
\email{simon.vanlangendonck@kuleuven.be}
\author{N.~Severijns}
\affiliation{KU Leuven, Instituut voor Kern- en Stralingsfysica, Celestijnenlaan 200D, B-3001 Leuven, Belgium}
\author{L.~Hayen}
\affiliation{Department of Physics, North Carolina State University, Raleigh, 27607 North Carolina, USA}
\affiliation{Triangle Universities Nuclear Laboratory, Durham, 27710 North Carolina, USA}
\author{F.~Gl\"uck}
\affiliation{Institute for Astroparticle Physics (IAP), Karlsruhe Institute of Technology (KIT), Hermann-von-Helmholtz-Platz 1, 76344 Eggenstein-Leopoldshafen, Germany}

\date{\normalsize\today} 

\begin{abstract}
Measurements of the beta decay correlation coefficients in nuclear decay aim for a precision below $1\%$ and theoretical predictions should follow this trend. In this work, the influence of the two dominant Standard Model correction terms, i.e. the recoil-order and the radiative correction, are studied for the most commonly measured beta correlations, i.e. the $\beta$-asymmetry parameter ($A_{\beta}$) and the $\beta-\nu$ angular correlation ($a_{\beta \nu}$). The recoil-order correction is calculated with the well-known Holstein formalism using the impulse approximation to evaluate experimentally inaccessible form factors. For the $\beta-\nu$ angular correlation previously unpublished, semi-analytical radiative correction values are tabulated.
Results are presented for the mirror beta decays up to $A=45$.
We examine the effect of both corrections and provide a comparison between different isotopes. This comparison will help planning, analysing, and comparing future experimental efforts.

\end{abstract}

\maketitle 

\section{Introduction}
Precision measurements in nuclear and neutron decay have played a prominent role in the progress of particle physics \cite{Dubbers2011, Bhattacharya2012, Cirigliano2013, Vos2015, Gonzalez2019, Falkowski2021}. They provided, for example, the basis for developing the vector (V) and axial-vector (A) current-current interaction \cite{PhysRev116}, led to the discovery of the maximal parity violation of weak interactions \cite{PhysRev105}, and determine the value of the axial-vector coupling constant $g_A$ \cite{Markisch2019, Falkowski2021}. All this is embedded in the framework of the Standard Model of particle physics (SM), which provides a very powerful tool to describe nature at the smallest scales. Precision experiments are continuously ongoing to shed light on some of the remaining mysteries, such as the determination of $V_{ud}$, the \textit{up-down} quark mixing matrix element, or to search for an exotic, potentially non V-A, current. These exotic currents are possible extensions of the electroweak interaction included in the SM. Bounds on the exotic currents are obtained within an effective field theory framework and strengthened by the model-independent analyses of the data \cite{Gonzalez2019, Falkowski2021}. In addition to other observables, the $\beta$ decay correlation coefficients provide a window on these exotic currents.
In the present work, we focus on such correlation coefficients for the mirror beta transitions up to $A=45$. The mirror beta transitions are mixed Fermi/Gamow-Teller transitions between an isospin $T=1/2$ doublet. They are of special interest because of the high degree of theoretical control over the nuclear matrix elements \cite{Falkowski2021, Severijns2008, VudNaviliatSeverijns, new19Ne}.
In addition to their intrinsic sensitivity to new physics, correlation measurements allow, in combination with the $\mathcal{F}$t-values of the superallowed Fermi decays, to determine the $V_{ud}$ quark mixing matrix element \cite{Falkowski2021, VudNaviliatSeverijns, Hayen2021}. 
An overview of the characteristics of the mirror $\beta$ transitions, i.e. nuclear spin, half-life and the Gamow-Teller to Fermi mixing ratio in the beta decay, as well as previous or ongoing experimental efforts is included in Table \ref{tab:meas}. Apart from tritium, $^{3}$H, all listed decays are $\beta^+$. 

In searching for exotic scalar or tensor type contributions to the weak interaction, experimental results are compared to SM calculations, in which a discrepancy might imply a sign of new physics. To guarantee a correct comparison and to avoid systematic errors, all necessary corrections should be included.
For neutron decay, which is the lightest mirror beta decay, the standard model correlation coefficients have already been calculated to first order in $1/M$ and $\alpha/\pi$ \cite{Wilkinson1982, Gardner2001, Ando2004, Ivanov2013}. 
A recent publication \cite{Hayen_tbp} presented a renewed, consistent description of the correlation coefficients, with nuclear structure corrections calculated in the Behrens-B\"uhring formalism \cite{behrens1982}, while also focusing on the dependence upon the experimental geometry. Qualitative results are discussed for the light mirror nuclei, for which ab initio nuclear calculations are within reach. The present article expands the discussion up to $^{45}$V, evaluating the nuclear structure effects (within the Holstein formalism \cite{Holstein1974}) as well as the radiative correction, and outlines their influence. This evaluation is motivated by a recent review on the largest contribution to the recoil-order corrections, i.e. the weak magnetism form factor \cite{WM_tbp}. 
Usually, the relevant corrections are only calculated for a single isotope while analysing the available experimental data. Our study examines the effect of the corrections more broadly and provides a comparison between different isotopes. This provides important information for planning, analysing, and comparing future experimental efforts.

Rather than provide a new formalism, this paper will carefully evaluate and analyse existing results and their implications. In Sec. \ref{sec:correlations}, the relevant results for the recoil-order and radiative corrections are summarised. Next, the paper focuses on the beta-asymmetry parameter in Sec. \ref{sec:asymmetry}. The size of both corrections is described, and their effect in an experimental analysis is estimated. Finally, Sec. \ref{sec:betaneut} considers the beta-neutrino correlation. Again, the size of the corrections for this correlation is described and previously unpublished, semi-analytical radiative correction values are tabulated.

\begin{table}
\caption{Spin, $J$, half-life, $T_{1/2}$, Gamow-Teller to Fermi mixing ratio, $\rho$ \cite{WM_tbp}, as well as existing or proposed measurements of correlation coefficients for the mirror beta decays up to A = 45.}
\label{tab:meas}
\begin{tabular}{c| cccc}
\hline \hline
Parent & $J$ & $T_{1/2} [s]$ & $\rho$ & Measurement  \\ \hline
$^{3}$H   & $1/2$ & $38854(35) \times 10^{4}$ & -2.1053(14) & \\
$^{11}$C  & $3/2$ & $1220.41(32)$ & -0.7544(8) & $a_{\beta \nu}$ \cite{Twinsol2016} \\
$^{13}$N  & $1/2$ & $597.88(23)$  & -0.5596(14) & $a_{\beta \nu}$ \cite{Twinsol2016} \\
$^{15}$O  & $1/2$ & $122.27(06)$  & 0.6302(16)  & $a_{\beta \nu}$ \cite{Twinsol2016} \\
$^{17}$F  & $5/2$ & $64.366(26)$  & 1.2955(10)  & $a_{\beta \nu}$ \cite{Twinsol2016}, $A_{\beta}$ \cite{17F} \\
$^{19}$Ne & $1/2$ & $17.2573(34)$ & -1.6020(9) & $a_{\beta \nu}$ \cite{PhysRev116}\\
 & & & & $A_{\beta}$ \cite{19Ne, new19Ne, Lienard2015} \\
$^{21}$Na & $3/2$ & $22.4527(67)$ & 0.7125(12)  & $a_{\beta \nu}$ \cite{21Na, 21Na2} \\
$^{23}$Mg & $3/2$ & $11.3050(44)$ & -0.554(2) & \\
$^{25}$Al & $5/2$ & $7.1674(44)$  & 0.8084(11)  & \\
$^{27}$Si & $5/2$ & $4.1112(18)$  & -0.6966(9) & \\
$^{29}$P  & $1/2$ & $4.1031(58)$  & 0.538(2)  & $A_{\beta}$ \cite{29P}\\
$^{31}$S  & $1/2$ & $2.5539(23)$  & -0.5294(15) & \\
$^{33}$Cl & $3/2$ & $2.5059(25)$  & -0.314(3) & \\ 
$^{35}$Ar & $3/2$ & $1.7752(10)$  & 0.282(2)  & $A_{\beta}$ \cite{35Ar, 35Ar2, 35Ar3, Lienard2015} \\
$^{37}$K  & $3/2$ & $1.23634(76)$ & -0.5779(15) & $A_{\beta}$\cite{37K}, $B_{\nu}$ \cite{nu_assym} \\
$^{39}$Ca & $3/2$ & $0.86046(80)$ & 0.6606(16)   & \\
$^{41}$Sc & $7/2$ & $0.5962(22)$  & 1.074(4)   & \\
$^{43}$Ti & $7/2$ & $0.5223(57)$  & -0.810(17)  & \\
$^{45}$V  & $7/2$ & $0.5465(51)$  & 0.64(2)   & \\
\end{tabular}
\end{table}

\section{Correlation coefficients} \label{sec:correlations}

The decay rate distribution $d\Gamma$ of an allowed beta decay is expressed in terms of the electron and neutrino momentum, electron energy, and the spin of the decaying system via \cite{Jackson1957, Hayen2018},
\begin{equation} \label{eq:correlations}
    \begin{aligned}
    d\Gamma = {} 
    & d\Gamma_0 \xi \bigg\{ 1+a_{\beta\nu}\frac{\vec{p} \cdot \vec{p}_{\nu}}{W W_{\nu}} + b_{F} \frac{1}{W} \\
    & + \frac{\langle \vec{J} \rangle}{J} \big[ A_{\beta} \frac{\vec{p}}{W} + B_{\nu} \frac{\vec{p}_\nu}{W_{\nu}} + D_{\times} \frac{\vec{p} \times \vec{p}_{\nu}}{W W_{\nu}} \big] \bigg\} 
    \end{aligned}
\end{equation}
in beta decay units of $\hbar\!=\!c\!=\!m_e\!=1$. 
Here, $W = E/m_ec^2 + 1$ is the total energy in units of the electron rest mass, $E$ being the kinetic energy, $p=\sqrt{W^2-1}$ the momentum of the electron or the neutrino (with subscript $\nu$), $\vec{J}$ is the initial nuclear polarisation vector and
\begin{equation}
    d\Gamma_0 = \frac{G^2_F V^2_{ud}}{(2 \pi)^5} C(W, W_0, Z) dW d\Omega d\Omega_{\nu}
\end{equation}
with $\Omega$ the angular coordinates, and $G_F$ the Fermi coupling constant. 
The factor $C(W, W_0, Z)$ is given by 
\begin{equation}
C(W, W_0, Z) = F_0L_0C_{sh}K(W, W_0,Z)p W (W_{0} - W)^2
\end{equation}
and consists of a combination of the Fermi function $F_0L_0$, the spin-independent shape factor $C_{sh}$, the phase space factor for allowed $\beta$ decay $pW(W_0 - W)^2$, and finally $K(W, W_0,Z)$ corresponding to higher-order corrections of varying nature, all of these being discussed for the $\beta$ spectrum shape in detail in Ref. \cite{Hayen2018}. 

The prefactor $\xi$ in Eq. \eqref{eq:correlations} incorporates the Fermi and Gamow-Teller nuclear matrix elements, $M_F$ and $M_{GT}$, and the beta decay coupling constants \cite{Jackson1957}. The ratio between the Gamow-Teller to Fermi strength ratio in the beta decay is defined by $\rho \simeq g_A M_{GT}/g_V M_F$. SM values for $\rho$ can be obtained from the comparative half-life, i.e. the corrected $\mathcal{F}t$-value, of the $\beta$ decay \cite{Severijns2008, WM_tbp}. The determination of this corrected $\mathcal{F}t$-value starts from the 'normal' $ft$-value, i.e. the product of the statistical rate function, $f$, and the partial half-life of the decay, $t$. Its experimental determination requires measuring three observables, i.e. the half-life of the decaying state, and the branching ratio and the $Q$-value of the transition. Correcting the $ft$-value with the appropriate radiative and isospin-symmetry breaking corrections the $\mathcal{F}t$-values are obtained \cite{Severijns2008}. Values for $\rho$ as presented in Ref. \cite{WM_tbp}, obtained by using the latest experimental results, are found in Table \Ref{tab:meas} where the sign of $g_A$ is defined positive as in the case of \cite{Holstein1974}. This renewed evaluation resolves a partial double counting in the electroweak renormalisation. The overall result are small shifts with respect to the earlier work of Ref. \cite{Severijns2008}, except for $^3$H which now has a different sign. 

The parameters, $a_{\beta \nu}$, $b_F$, $A_{\beta}$, $B_{\nu}$, and $D_{\times}$ are the so-called correlation coefficients, which represent the amplitude in the different correlations, and are the observables in dedicated experiments. If these correlations are measured sufficiently precise, a comparison with the expected value as calculated from the SM provides information on the size or absence of possible exotic interactions. 

\subsection{Asymmetry parameter and $\beta \nu$-correlation}

\begin{figure*}
	\centering
	\includegraphics[width=\textwidth]{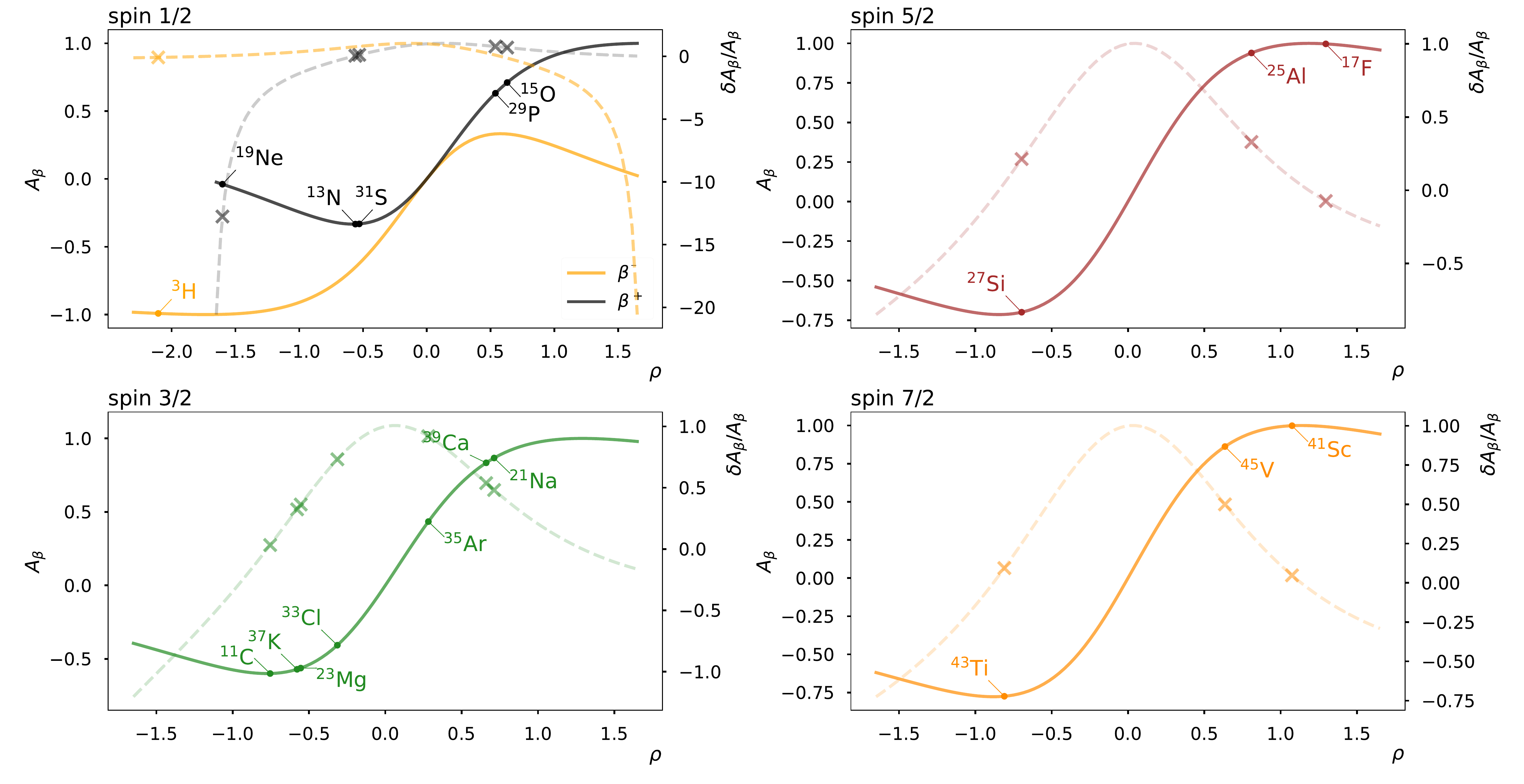}
	\caption{Size (full line) and sensitivity (dashed line, note the different y-axis) of the beta-asymmetry parameter $A_{\beta}$ with respect to $\rho$ for different nuclear spins. The discussed mirror transitions are highlighted. For $J=1/2$, the cancellation of the first order expression in Eq. \eqref{eq:A_0} at $\rho=\pm\sqrt{3}$ is clearly visible.}
	\label{fig:A_sensi}
\end{figure*}

The leading-order standard model expression for the beta-asymmetry parameter, $A_{\beta,0}$, i.e. omitting recoil-order and radiative corrections, is energy independent and given by \cite{Severijns2008}
\begin{equation} \label{eq:A_0}
  A_{\beta,0} = \frac{\mp \Lambda_{J_i,J_f}\rho^2 + 2 \delta_{J_i J_f} \sqrt{\frac{J_i}{J_i+1}}\rho}{1+\rho^2}
\end{equation}
where the upper (lower) sign is for $\beta^-$ ($\beta^+$) decay and $J_{i,f}$ is the spin of the mother and daughter nuclear states, respectively. For the mirror $\beta$ transitions, with $J_i = J_f$, one has $\Lambda_{J_i,J_f} = 1 / (J + 1)$ \cite{Jackson1957}. Note that the sign of the second term in the denominator depends on the sign used for $g_A$, which is positive here following Ref. \cite{Holstein1974}.
The size of the beta-asymmetry parameter $A_{\beta,0}$ with respect to $\rho$ is illustrated in Fig. \ref{fig:A_sensi} for different $J$ values. Table \ref{tab:sensitivity} lists its size for the different mirror transitions under consideration. Due to the sign change for $\rho$ in $^{3}$H \cite{WM_tbp}, the value for $A_{\beta,0}$ changes drastically from -0.09408(46) \cite{Severijns2008} to -0.99145(5).

\begin{table}
\caption{Values for the leading-order SM value and its sensitivity to $\rho$ of the beta-asymmetry parameter, $A_{\beta}$, and the beta-neutrino angular correlation, $a_{\beta \nu}$.}
\label{tab:sensitivity}
\begin{tabular}{c| cccc}
\hline \hline
  Parent &
  $A_{\beta, 0}$&
  $\delta A_{\beta}/A_{\beta}$&
  $a_{\beta \nu,0}$ & 
  $\delta a_{\beta \nu}/a_{\beta \nu}$  \\ \hline
$^{3}$H   & -0.99145(5)  & -0.08  & -0.0879(3)   &  4.56 \\
$^{11}$C  & -0.59974(2)  & 0.03   &  0.5164(6)   & -1.19 \\
$^{13}$N  & -0.33309(4)  & 0.05   &  0.6820(12)  & -0.71 \\
$^{15}$O  & 0.7103(13)   & 0.70   &  0.6210(14)  & -0.87 \\
$^{17}$F  & 0.99664(6)   & -0.07  &  0.1645(5)   & -3.79 \\
$^{19}$Ne & -0.0389(3)   & -12.8  &  0.0405(3)   & -13.3 \\
$^{21}$Na & 0.8668(7)    & 0.48   &  0.5510(10)  & -1.08 \\
$^{23}$Mg & -0.5628(7)   & 0.36   &  0.6868(17)  & -0.70 \\
$^{25}$Al & 0.9393(4)    & 0.33   &  0.4730(9)   & -1.35 \\
$^{27}$Si & -0.6994(2)   & 0.21   &  0.5644(8)   & -1.04 \\
$^{29}$P  & 0.6314(19)   & 0.79   &  0.7007(18)  & -0.66 \\
$^{31}$S  & -0.33154(11) & 0.12   &  0.7081(13)  & -0.64 \\
$^{33}$Cl & -0.407(3)    & 0.73   &  0.880(2)    & -0.25 \\
$^{35}$Ar & 0.4342(3)    & 0.92   &  0.9018(15)  & -0.20 \\
$^{37}$K  & -0.5710(5)   & 0.32   &  0.6662(13)  & -0.75 \\
$^{39}$Ca & 0.8340(11)   & 0.54   &  0.5949(14)  & -0.95 \\
$^{41}$Sc & 0.99872(17)  & 0.05   &  0.286(2)    & -2.32 \\
$^{43}$Ti & -0.7747(15)  & 0.09   &  0.472(13)   & -1.35 \\
$^{45}$V  & 0.862(14)    & 0.50   &  0.617(17)   & -0.88 \\
\end{tabular}
\end{table}

\begin{figure}
	\centering
	\includegraphics[width=\columnwidth]{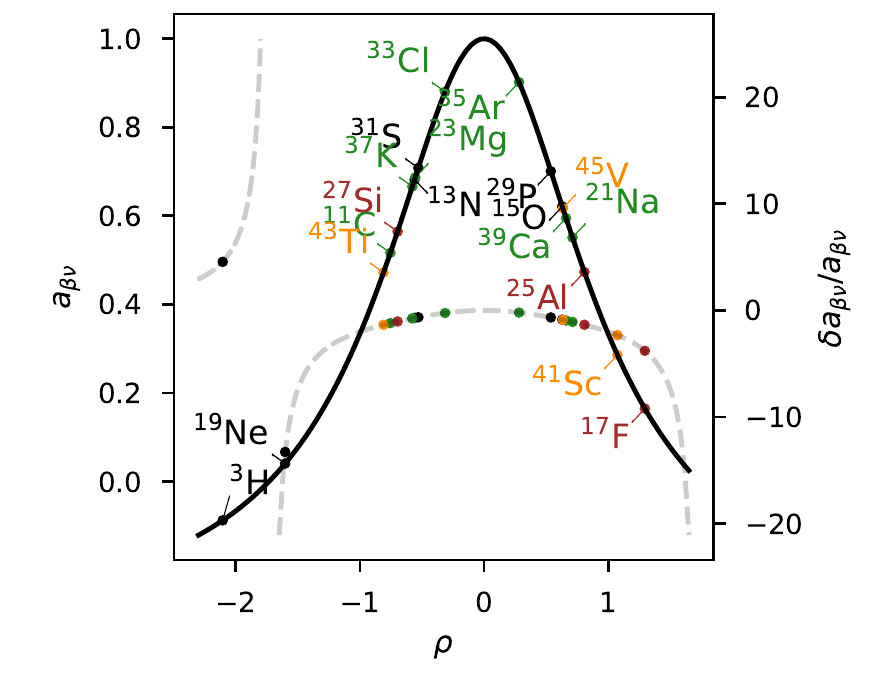}
	\caption{Size (full line) and sensitivity (dashed line, note the different y-axis) of the beta-neutrino correlation $a_{\beta \nu}$ with respect to $\rho$, with colours corresponding to the nuclear spin. Sensitivity enhancements are obtained for $\rho=\pm\sqrt{3}$.}
	\label{fig:abnu_sensi}
\end{figure}

The leading-order Standard Model expression for the beta-neutrino angular correlation, $a_{\beta \nu, 0}$ of the mirror nuclei, omitting recoil-order and radiative corrections, is given by \cite{Severijns2008}:
\begin{equation} \label{eq:a_sm}
	a_{\beta \nu,0} = \frac{1 - \rho^2/3}{1+\rho^2}.
\end{equation}
Like the beta-asymmetry parameter, this expression is energy independent. Values for the beta-neutrino correlation $a_{\beta \nu}$ with respect to $\rho$ are illustrated in Fig. \ref{fig:abnu_sensi} and its size for the different mirror transitions again given in Table \ref{tab:sensitivity}.

\subsection{Sensitivity $A_{\beta}$ and $a_{\beta \nu}$ to $\rho$} \label{sec:sensitivity}

At present, not more than a handful beta-asymmetry measurements have been performed on the mirror nuclei (see Table \ref{tab:meas}), with only more recent studies \cite{37K} reaching sub-percent precision. However, when evaluating the impact of experimental results precision should be compared to the sensitivity of the isotope under consideration. 

The sensitivity to $\rho$ for the beta-asymmetry parameter is given by,
\begin{equation} \label{eq:A_sensi}
  \frac{\delta A_{\beta}}{A_{\beta}} = 2\frac{\mp \rho + \kappa(1-\rho^2)}{(1+\rho^2)(\mp \rho+2\kappa)} \frac{\delta \rho}{\rho}
\end{equation}
with $\kappa= \sqrt{J(J+1)}$.
In general, the cases with the highest sensitivity are located in the steeper parts of the curves in Fig. \ref{fig:A_sensi}. High sensitivity is obtained when $\rho$ is close to the first-order cancellation, as for $^{19}$Ne. Diverging sensitivity enhancement, and corresponding first-order cancellation, is obtained when $\rho=\pm 2\kappa$, i.e. $\rho=\pm\sqrt{3}\approx\pm1.73$, $\rho=\pm\sqrt{15}\approx\pm3.87$, $\rho=\pm\sqrt{35}\approx\pm5.92$, and $\rho=\pm\sqrt{63}\approx\pm7.94$ for $J=1/2$, $J=3/2$, $J=5/2$, and $J=7/2$, respectively. Except for $J=1/2$, these values lie far from the $\rho$ values obtained in the studied mirror $\beta$ transitions (see Table \ref{tab:meas}). Equation \eqref{eq:A_sensi} is illustrated in Fig. \ref{fig:A_sensi} (note the second y-axis) for the different isotopes under consideration with numerical values being tabulated in Table \ref{tab:sensitivity}. It is noteworthy that with the updated sign of $\rho$ for $^{3}$H  \cite{WM_tbp}, the sensitivity plummeted from $\delta A_{\beta}/A_{\beta}=5.01$ to $\delta A_{\beta}/A_{\beta}=-0.08$.

The sensitivity for the beta-neutrino correlation $a_{\beta \nu}$ is given by,  
\begin{equation}
  \frac{\delta a_{\beta \nu}}{a_{\beta \nu}} = -\frac{8\rho^2}{3(1+\rho^2)(1-\rho^2/3)} \frac{\delta \rho}{\rho}
\end{equation}
and is illustrated in Fig. \ref{fig:abnu_sensi}. A diverging sensitivity enhancement is obtained for $\rho=\pm\sqrt{3}$.

Clearly the more sensitive cases to extract $\rho$ from a measurement of $A_{\beta}$, and so determine $V_{ud}$ with a good precision, requiring a relative precision of typically 0.5\%, are $^{15}$O, $^{19}$Ne, $^{29}$P, $^{33}$Cl and $^{35}$Ar. For $a_{\beta \nu}$ the sensitivity is seen to be on average higher, with the better cases now being $^{3}$H, $^{17}$F, $^{19}$Ne and $^{41}$Sc (see also \cite{Hayen_tbp} and \cite{Severijns2013}).
However, when a measurement of $a_{\beta \nu}$ probes the recoil spectrum, either through direct measurement or via beta-delayed emission, the spectral distortion is proportional to $a_{\beta\nu}$. Therefore, the experimental sensitivity is proportional to the product of $\delta a_{\beta \nu}/a_{\beta\nu}$ and $a_{\beta\nu}$, both given in Table \ref{tab:sensitivity}. The diverging sensitivity enhancement disappears and unlike for $A_{\beta}$ measurements it cannot be recovered using subtractions or super ratios. Therefore, the figure of merit becomes more homogeneous for $a_{\beta \nu}$ measurements probing the recoil spectrum. The sensitivity can be recovered once more when, together with a recoil detection, an asymmetry is experimentally constructed through coincidence detection with the outgoing $\beta$ particle for a slice of phase space (e.g., parallel versus anti-parallel).

For both the beta-asymmetry parameter and the beta-neutrino correlation, $\rho$-values with a sensitivity enhancement coincide with small results in the leading-order expressions of Eq. \eqref{eq:A_0} and Eq. \eqref{eq:a_sm}. Sensitivity enhancements, thus, result in an increased importance of higher-order correction terms. The next paragraphs describe a formalism for the inclusion of two important and necessary higher order terms in high-precision and/or high-sensitivity measurements.

\subsection{Recoil effect} \label{sec:intro_rec}
Beta decay occurs within a nucleus and is affected by the surrounding nuclear environment and by the strong interaction. The so-called recoil terms induced by the strong interaction are as such an intrinsic part of precision experiments in nuclear beta decay. They are called recoil terms because they are of order $q/M$, $q^2/M^2$, ... , where $q = p_i-p_f$ is the momentum transfer and $M$ the average mass of the mother and daughter nucleus. 
Several useful formalisms exist to incorporate their effect. In this work the standard, well-known study by Holstein \cite{Holstein1974} will be used. Although the formalism has disadvantages \cite{Hayen_tbp}, most notably its incompatibility for forbidden decays and the post-hoc addition of electromagnetic effects, its easy symmetry properties prove very convenient when working with mirror beta decays. 

In the Holstein formalism the nuclear structure aspects are encoded into ten form factors, noted as $a$, $b$, $c$, $d$, $e$, $f$, $g$, $h$, $j_2$ and $j_3$. These allow for a model-independent analysis of the beta decay observables but can also be linked to the coupling constants and nuclear one-body matrix elements using, e.g., the impulse approximation. The form factors are $q^2$-dependent but in general this dependence is only retained for the dominant Fermi ($a$) and Gamow-Teller ($c$) form factors. These are then expanded as \citep{Holstein1974}
\begin{subequations}
\begin{align}
a(q^2) &\approx a_1 + a_2 \left(\frac{q}{M}\right)^2 \\
c(q^2) &\approx c_1 + c_2 \left(\frac{q}{M}\right)^2
\label{eq:c_expansion}
\end{align}
\label{eq:expansion}
\end{subequations}
while the others are approximated by their value for $q^2 = 0$. This approximation is possible due to the small momentum transfer in $\beta$ decay, i.e. $qR \ll 1$, with $R$ the nuclear radius \cite{WM_tbp}. As such, all nuclear structure information is incorporated in twelve parameters. \\
Because mirror decays are so-called analog decays, i.e. between states of the same isotopic multiplet, $d$, $e$, $f$ and $j_2$ involve pure second class currents \cite{Holstein1974}. Second-class currents transform differently than the leading term in the vector and axial vector current under $G$-parity, which is a charge conjugation followed by a 180$\degree$ rotation around the y-axis in isospin space $(G = C e^{i \pi T_2})$, and are absent in the SM (neglecting EM effects). To date, independent measurements found no evidence for their existence \cite{Holstein2014}. 

The calculation of the recoil-order corrections in the Holstein formalism for mirror decays involves an evaluation of the eight remaining and generally non-zero form factors. The ratio between the leading order form factors, $\rho=c/a$, and the largest second order contribution, the weak magnetism form factor $b \equiv b_{W\!M}$, can be independently determined \cite{WM_tbp}.
All other form factors are not experimentally accessible, and calculations such as those within the impulse approximation are necessary. For the spin $J = 1/2$ mirror decays, however, the calculations simplify because the form factors involving higher spin changes, i.e. the rank 2 form factors, $f$, $g$, and $j_3$, are zero. 

All non-leading form factors, except for $a_2$, were calculated in Ref. \cite{WM_tbp}. In that study, the calculated values were compared to the weak magnetism form factors obtained from available experimental data and found to reproduce the trends within the data. 
For this reason, we deem it appropriate to use these values for our order of magnitude estimation. Using these values, the second order correction to the Fermi decay $a_2$ remains the only undetermined form factor. However, its effect was estimated previously \cite{a2_estimate}. Assuming a uniform weak charge distribution and the conserved vector current hypothesis, the ratio between the leading and non-leading term of the dominant form factor is given by 
\begin{equation} \label{eq:a1a2}
\frac{a_2}{a_1} = \frac{R^2}{10}.
\end{equation} 
Note, however, that this expression is obtained using an expansion in $q^2$ instead of $q^2/M^2$ (as was presented in Eq. \eqref{eq:expansion}) and as such incorporates an additional factor $M^{-2}$. Also, the results for $c_2$, determined to be $\mathcal{O}(1 f\!m^2)$ for the here-discussed mirrors \cite{WM_tbp}, have been obtained using an expansion in $q^2$. Notice that both $a_2$ and $c_2$ have dimension $f\!m^2$, therefore, a factor of $(\hbar c)^2$ has to be added when evaluating their size with respect to $a_1$ or $c_1$, respectively. Due to this factor, it is possible to use $\rho \approx c_1/a_1$. 

In the impulse approximation, used to derive expressions for the experimentally inaccessible form factors, the second class form factors, $d$, $e$, $f$ and $j_2$, contain first class contributions \cite{Holstein1974}. These first class contributions result in possible non-zero values for these form factors. However, as mentioned in Ref. \cite{WM_tbp}, the first class contributions to $d$ and $j_2$ can, for the mirror nuclei, be shown to vanish or to be negligible small, respectively. 
For $f$ the dominant term vanishes only in the limit of isospin invariance, resulting in non-zero calculated values in the impulse approximation \cite{WM_tbp}. 
For the $e$ form factor a significant first class current contribution is apparently predicted by the naive impulse approximation due to the fact that the vector current is no longer divergenceless  \cite{Holstein1974}. In the original work \cite{Holstein1974}, this is regarded as a shortcoming of the impulse approximation at recoil level and as a reason that experiments should be analysed in a model independent way.

\subsection{Radiative correction} \label{sec:intro_rc}
The long history of calculations of electroweak radiative corrections is discussed by Sirlin \& Ferroglia \cite{Sirlin2013}. Due to the vastness of the subject, it is impossible to be all-encompassing, so although several observables are explicitly mentioned in Ref. \cite{Sirlin2013} correlation coefficients are not. The literature on the radiative correction to the correlation coefficients can roughly be divided into two parts and consists of several older, initial calculations by, for example, R.T. Shann \cite{Shann1971}, Yokoo \& Morita \cite{YM1973, YM1976}, and Garc\'{\i}a \& Maya \cite{GM1978}, and a later prolonged effort by F. Gl\"uck and collaborators \cite{Gluck1990, Gluck1992, Gluck1993, Gluck1997, Gluck1998, Gluck_tbp}. \\
Most calculations separate inner and outer radiative corrections as proposed by A. Sirlin \cite{Sirlin1967}, which are the model-dependent and model-independent terms, respectively. This separation splits the radiative correction into infrared divergent and convergent terms in a gauge-invariant way. Due to the large energy difference between the typical beta decay energy, i.e. a few MeV, and the scale governing the strong interaction, i.e. the pion mass $M_{\pi} \approx 140$ MeV, remaining energy-dependent terms in the inner correction are of order $\mathcal{O}(\alpha E/M_{\pi}) \sim 10^{-5}$, and hence can safely be neglected. 
A recent investigation \cite{Gorchtein2019} has shown new $\mathcal{O}(10^{-4})$ corrections to the spectral shape due to inelastic corrections in nuclei at the MeV scale. Its effect on correlation coefficients is at this moment not yet studied. Assuming these corrections to be of similar magnitude, we neglect these for now. \\
If the energy-dependent terms are omitted, the inner radiative correction, usually denoted as $\Delta_R$, can be absorbed by a renormalization of the effective coupling constants, i.e. $g_V \equiv g_{V,0}(1 + \Delta^V_R)$ and $g_A \equiv g_{A,0}(1 + \Delta^A_R)$. This renormalization was crucial to establish the universality of the electroweak interaction. The most recent calculated values of the inner radiative correction are found in Refs. \cite{Seng2018, Czarnecki2019, Hayen2021}. A discussion on its influence in the analysis of the $\mathcal{F}$t-values in superallowed $0^+ \rightarrow 0^+$ Fermi transitions can be found in Ref. \cite{HT2020}. \\
The earlier works with analytical expressions for radiative corrections on the correlation coefficients \cite{Shann1971, YM1973, YM1976, GM1978} do not always implement the proposed separation between model-dependent and model-independent terms \cite{Sirlin1967}. However, a comparison to earlier work for the $\beta$ spectrum shape \cite{Sirlin1959}, which has used $\rho = 1$, shows their agreement. In the present work, only the energy-dependent outer correction is of interest.
The outer correction concerns both diagrams with virtual photon exchange and with real photons in the final state, so-called radiative beta decay. Both types of diagram give an infrared infinite correction which is resolved by combining them with the electron wave function renormalization. Only the combined result becomes finite and gauge-invariant.
Rewriting the expressions from the original work gives \cite{Shann1971, YM1973},
\begin{equation} \label{eq:N_asymmetry}
    \begin{aligned}
    d\Gamma = {} 
    & d\Gamma_0 \xi \bigg[ \left( 1 + \frac{\alpha}{2\pi} g(\beta) \right) \\
    & - P\frac{p}{W} \cos{\theta} A_{\beta,0} \left( 1 + \frac{\alpha}{2\pi} h(\beta) \right) \bigg], 
    \end{aligned}
\end{equation}
with $|M_F|$ the Fermi matrix element, $P$ the nuclear polarisation, $\beta=p/W=v/c$ is the velocity of the beta particle, and $\theta$ the angle between $\vec{P}$ and the emitted electron.
The two outer correction functions, $g(\beta)$ and $h(\beta)$, are given by the well-known general radiative correction function as defined by Sirlin \cite{Sirlin1967},
\begin{equation} \label{eq:gSirlin}
\begin{aligned}
g\left(W_{0}, W\right)& = 3 \ln \left(m_{p}\right)-\frac{3}{4}+\frac{4}{\beta} L_{s}\left(\frac{2 \beta}{1+\beta}\right) \\
  & + 4 \left(\frac{\tanh ^{-1} \beta}{\beta}-1\right) \bigg[\frac{W_{0}-W}{3 W}-\frac{3}{2} \\
  & + \ln \left[2\left(W_{0}-W\right)\right]\bigg] + \frac{\tanh ^{-1} \beta}{\beta} \bigg[2\left(1+\beta^{2}\right) \\
  & +\frac{\left(W_{0}-W\right)^{2}}{6 W^{2}} - 4 \tanh ^{-1} \beta \bigg]
\end{aligned}
\end{equation}
and by Shann \cite{Shann1971}
\begin{equation} \label{eq:h}
\begin{aligned}
h\left(W_{0}, W\right)& = 3 \ln \left(m_{p}\right)-\frac{3}{4}+\frac{4}{\beta} L_{s}\left(\frac{2 \beta}{1+\beta}\right) \\
  & + 4\left(\frac{\tanh ^{-1} \beta}{\beta}-1\right) \bigg[\frac{W_{0}-W}{3 W\beta^2}-\frac{3}{2} \\
  & + \ln \left[2\left(W_{0}-W\right) \right] + \frac{(W_{0}-W)^2}{24 W^2\beta^2} \bigg] \\
  & + \frac{4\tanh ^{-1} \beta}{\beta}\left[1 - \tanh ^{-1} \beta\right],
\end{aligned}
\end{equation}
with $m_p$ the mass of the proton, and the here introduced Spence function $L_{s}(x)$ given by,
\begin{equation} \label{eq:spence}
    L_s(x) = \int_0^x \frac{dt}{t} \ln (1-t).
\end{equation} 
For the beta-asymmetry parameter the analytical formula presented in this section can be used. However, when the recoiling nucleus is detected in an electron-neutrino angular correlation measurement it will no longer be valid as will be explained in Sec. \ref{sec:betaneut}. In what follows the impact of both recoil-order and radiative corrections is discussed for both the beta-asymmetry and the beta-neutrino correlation. 

\section{Beta-asymmetry} \label{sec:asymmetry}
The most straightforward way to determine the beta asymmetry is by measuring the difference between the number of events in up- and downstream detectors with respect to the nuclear polarisation. In practice, some more complicated super-ratio might be used which allows a cancellation of systematic errors \cite{new19Ne, 37K, superratio}.
The obtained asymmetry is energy dependent and related to the beta-asymmetry parameter $A_{\beta}$ via 
\begin{equation} \label{eq:A_exp}
A_{\beta}(W) = \langle \cos \theta \rangle \beta A_{\beta} P C
\end{equation}
with $\langle \cos \theta \rangle$ the average value of the cosine of the solid angle of the detector geometry, $P$ the polarisation of the mother nucleus and $C$ (potential) correction terms depending on the experimental configuration. \\
The recoil-order and radiative corrections will lead to an energy-dependent expression for the beta-asymmetry parameter $A_{\beta}$ on the right-hand side of Eq. \eqref{eq:A_exp}. If the corrections are neglected but are nevertheless sufficiently large in comparison to the experimental precision, a systematic deviation in the analysis of Eq. \eqref{eq:A_exp} will be introduced. 

\subsection{Recoil effect}
To incorporate the recoil effect, Eq. \eqref{eq:A_0} is, in the Holstein formalism, replaced by 
\begin{equation} \label{eq:A_rec}
	A_{\beta,R} = \frac{H_1 + \Delta F_4 + \Delta F_7/3}{H_0 + \Delta F_1} 
\end{equation}
with $H_i$ and $\Delta F_i$ spectral functions as defined in the appendices in \cite{Holstein1974}. Besides the recoil-order corrections, which are contained in $H_i$, this expression also contains a, post-hoc added, electromagnetic correction, $\Delta F_i$. This electromagnetic correction is due to the Coulomb field between the nucleus and the outgoing lepton.
\begin{widetext}
For the mirror decays we have, 
\begin{equation} \label{eq:H0}
 \begin{aligned}
    H_0(W) =&~1 + \frac{2W}{M} + \frac{2}{3} \frac{a_2}{a_1}  \left(1 + 4W W_0 + 2 \frac{W_0}{W} - 4W^2 \right) \\
    & + \frac{2}{9} \frac{c_2}{c_1} \left( 11 + 20 W W_0 - 2\frac{W_0}{W} - 20W^2 \right) \\
    &+ \left(\frac{c_1}{a_1}\right)^2 \bigg[ 1 - \frac{2W_0}{3M} \left(1 \pm \frac{b}{c_1} \right) + \frac{4W}{3M} \left( \frac{5}{2} \pm \frac{b}{c_1} \right) - \frac{2}{3MW} \left(1 \pm \frac{b}{c_1} - \frac{h}{c_1}\frac{W_0 - W}{4M} \right) \bigg] 
 \end{aligned}
\end{equation}
and
\begin{equation} \label{eq:H1}
\begin{aligned}
  H_1 (W, J) = & ~ 2 \frac{c_1}{a_1} \sqrt{\frac{J}{J+1}} \bigg[1 - \frac{W_0}{3M} \left(1 \pm \frac{b}{c_1} \right) + \frac{W}{3M} \left(7 \pm \frac{b}{c_1} \right) - 2 \left( \frac{c_2}{c_1} + \frac{a_2}{a_1} \right) \frac{4W(W_0-W) + 3}{3} \bigg] \\
  & + \frac{c_1}{a_1} \sqrt{\frac{3}{2}} \frac{g}{c_1} \frac{\sqrt{(2J-1)(2J+3)}}{J+1}   \frac{W_0^2 - 11 W W_0 + 6 + 4W^2}{6M^2} \\
  & \mp \left(\frac{c_1}{a_1}\right)^2 \frac{1}{J+1} \bigg[ 1 - \frac{2 W_0}{3M} \left(1 \pm \frac{b}{c_1} \right) + \frac{W}{3M} \left(11 \pm 5 \frac{b}{c_1} \right) + 2\frac{c_2}{c_1} \frac{8W(W_0-W)+3}{3} \bigg]     
\end{aligned}
\end{equation}
omitting form factors that are zero due to symmetry considerations. The upper (lower) sign refers to electron (positron) decay. In these equations, terms depending on $a_2$ or $c_2$ can differ by a factor $M^2$ from other works depending on whether an expansion in $q^2/M^2$ \cite{Holstein1974} or in $q^2$ \cite{a2_estimate, WM_tbp} is used. As discussed before, in the evaluation of these expressions we will use the approximation $c_1/a_1\approx\rho$.
\end{widetext}
The modifications of the spectral functions due to the Coulomb effect $\Delta F_i$ with $i=1,4,7$ are (Eq. C4 in Ref. \cite{Holstein1974}),
\begin{subequations} \label{eq:deltaFi}
\begin{align}
  \Delta F_1(W) = \mp \frac{8 \alpha Z}{ 3 \pi} & \bigg[ \bigg( 8 W + W_0 + \frac{3}{W} \bigg) \label{eq:deltaF1} \\ 
  & + \frac{\rho^{2}}{3} \bigg( 28W -W_0 +\frac{9}{W} \bigg) \bigg] X, \nonumber \\
  \Delta F_4(W, J)  = \mp \frac{8 \alpha Z}{ 3 \pi} & \bigg[ \sqrt{\frac{J}{J+1}} 2 \rho \mp \frac{\rho^2}{J+1} \bigg] 9WX, \label{eq:deltaF4}\\
  \Delta F_7(W, J)  = \mp \frac{8 \alpha Z}{ 3 \pi} & \bigg[ \sqrt{\frac{J}{J+1}} 2 \rho \label{eq:deltaF7} \pm \frac{\rho^2}{J+1} \bigg] (W_0 - W) X,  
\end{align}
\end{subequations}
where the value for $X$ depends on the used electric and weak charge density. Using a uniform charge density or a surface charge distribution results in $X=9\pi R/140$ or $X=\pi R/12$, respectively \cite{Holstein1974}. All results presented here will use the former value and will use $R=1.2A^{1/3}$ for the nuclear radius. The difference between both is small and will increase for higher $Z$ due to the prefactor, $8\alpha Z/3\pi$. The largest absolute difference in Eq. \eqref{eq:A_rec} (for $^{45}$V) amounts to $4 \times 10^{-4}$.

The incorporation of the electromagnetic correction has been mentioned before as one of the disadvantages of the Holstein formalism. In contrast to calculations in the Behrens-B\"uhring formalism, these corrections are not directly included and the expressions in Eq. \eqref{eq:deltaFi} are the result of separate calculations \cite{Calaprice1976, Holstein1974PRC}. In these calculations the ratio between the leading and non-leading term of the dominant form factors is approximated for both $a$ and $c$ using Eq. \eqref{eq:a1a2}, i.e. $c_2/c_1=a_2/a_1=R^2/10$. For the beta spectrum shape of pure Gamow-Teller decays, an expression reinstating the explicit $c_2/c_1$ dependence is given in Ref. \cite{Calaprice1976}. A similar expression can also be found for the Fermi part but is, to our knowledge, only published for the specific case of the recoil energy spectrum \cite{Kleppinger1977}. To evaluate the importance of this approximation, the impulse approximation results for $c_2/c_1$ \cite{WM_tbp} are compared to $R^2/10$. 
The former are found to be on average a factor $2.0$ larger than the $R^2/10$ values with a scatter of about 15\%.

\begin{figure}
	\centering
	\includegraphics[width=\columnwidth]{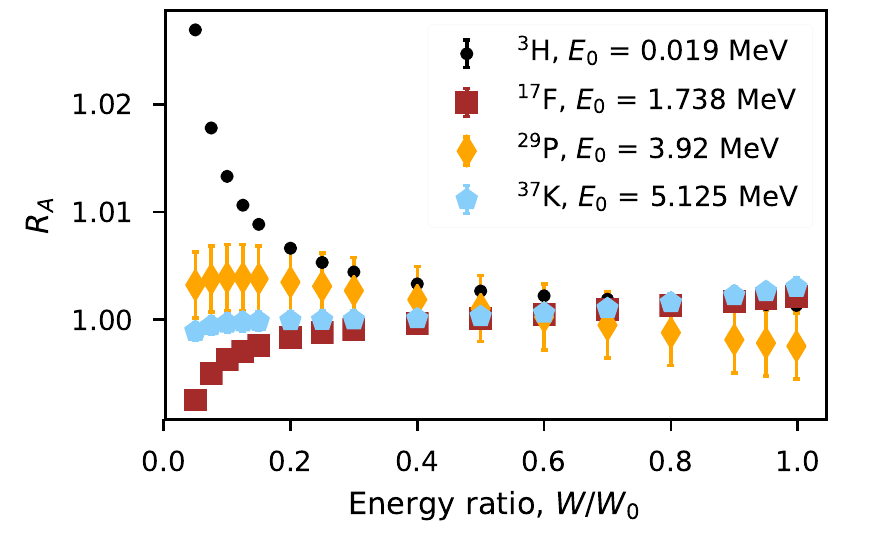}
	\caption{The influence of the recoil correction terms on $A_{\beta}$ for mirror nuclei with different endpoint energies. No clear trend is observed in the size of the correction terms with increasing endpoint energy. Energies are given as the ratio between the decay energy, $W$, and the endpoint energy, $W_0$. Error bars include the uncertainty from Ref. \cite{WM_tbp} on the end point energy, the weak magnetism form factor $b_{W\!M}$, and the Gamow-Teller to Fermi strength ratio $\rho$. The latter are the dominant contribution.}
	\label{fig:A_recoil}
\end{figure}

\begin{figure}
	\centering
	\includegraphics[width=\columnwidth]{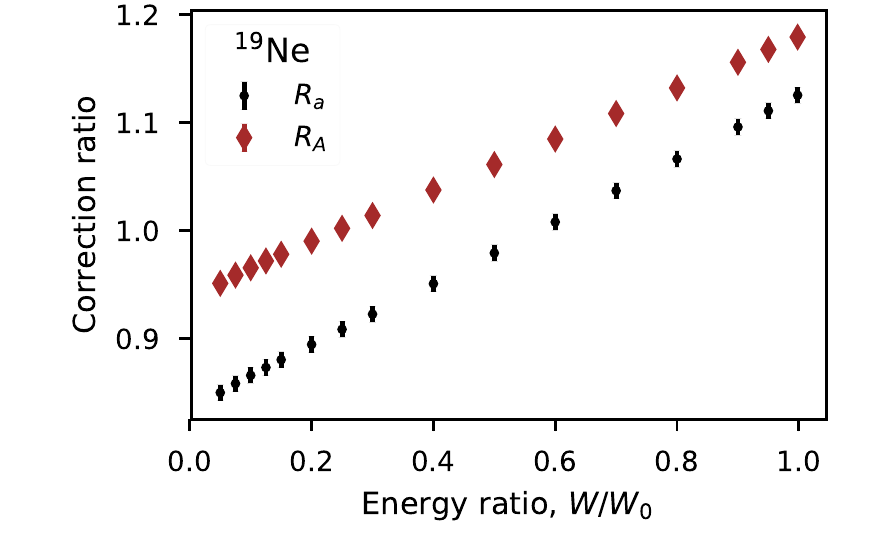}
	\caption{The influence of the recoil correction terms for $^{19}$Ne for both $A_{\beta}$ and $a_{\beta \nu}$. For this isotope, the leading order expressions are suppressed, therefore, the relative importance of the higher-order corrections increases drastically.}
	\label{fig:Ne_corr}
\end{figure}

To illustrate the effect of the recoil terms on the asymmetry in mirror beta decays we define a correction ratio $R_A$ as follows 
\begin{equation} \label{eq:R_A}
  R_A = \frac{A_{\beta,R}(W)}{A_{\beta,0}},
\end{equation}
which compares the size of the recoil-order correction with respect to the first order expression in Eq. \eqref{eq:A_0}. An identical expression will be used for the beta-neutrino correlation $a_{\beta \nu}$. This ratio provides an indication for the experimental precision at which recoil terms become important. Figure \ref{fig:A_recoil} shows this ratio for selected mirror beta decays with different endpoint energies using the form factors tabulated in Ref. \cite{WM_tbp}. The correction behaves linearly in the upper part of the energy spectrum with the total deviation limited to about a percent or less. Especially for the nuclei with low endpoint energy a deviation from this linear trend is seen at lower energies.
However, $^{19}$Ne is a notorious exception. Here, the first order expression is suppressed, such that the relative importance of the recoil corrections increases substantially. In this case, the impact of the recoil-order corrections is very important and crucial already at the percent-level precision \cite{19Ne, new19Ne}. The importance of the corrections is shown in Fig. \ref{fig:Ne_corr} together with the later discussed correction for the beta-neutrino correlation. 

The importance of the recoil-order correction is further inspected by performing an idealised experimental analysis assuming full polarisation $P = 1$, $\langle \cos \theta \rangle =1/2$, and absence of additional experimental corrections, i.e. $C = 1$. When Eq. \eqref{eq:A_exp} is used as fit function for the, energy-independent, asymmetry parameter $A_{\beta}$ from the experimentally determined function $A_{\beta}(W)$, any unaccounted for energy-dependent terms, such as the recoil effect, will shift the result away from $A_{\beta,0}$ (Eq. \eqref{eq:A_0}). 
The (normalised) shift of the fit result $A_{\beta, fit}$ from $A_{\beta,0}$ (Eq. \eqref{eq:A_0}) when fitting $A_{\beta}(W)$ (Eq. \eqref{eq:A_exp}) with inclusion of the correction under consideration, and assuming equal weights for all energies, is defined as
\begin{equation} \label{eq:deltaA}
 \Delta A_{\beta,i}=\frac{A_{\beta,0}-A_{\beta,fit}}{A_{\beta,0}}.   
\end{equation}
This is used as our estimate of the systematic error induced by neglecting the correction $i=$WM, R, RC with WM
= Weak Magnetism only, R = full Recoil correction, and RC =
Radiative Correction. The obtained results are shown in Table \ref{tab:offset}.

\begin{table}
\caption{Values for the systematic error in an idealised experimental analysis when the corresponding correction is neglected (WM = Weak Magnetism only, R = full Recoil correction, and RC = Radiative Correction) obtained using Eq. \eqref{eq:deltaA}. More details are given in the text.}
\label{tab:offset}
\begin{tabular}{c|rrr}
\hline \hline
\multicolumn{1}{c}{Parent}  & 
  \multicolumn{1}{c}{$\Delta A_{\beta, R}$} &
  \multicolumn{1}{c}{$\Delta A_{\beta, W\!M}$} &
  \multicolumn{1}{c}{$\Delta A_{\beta, RC}$ }  \\
\multicolumn{1}{c}{}  & 
\multicolumn{1}{c}{$[10^{-3}]$} &
\multicolumn{1}{c}{$[10^{-3}]$} &
\multicolumn{1}{r}{$[10^{-4}]$} \\\hline
$^{3}$H    & -3.0    & -3.1   & 22 \\
$^{11}$C   & -1.5   & -1.6 & -10  \\
$^{13}$N   & -1.5   & -1.6  & -8   \\
$^{15}$O   & 0.2   & -0.1  & -6   \\
$^{17}$F   & 0.3   & -0.3  & -6   \\
$^{19}$Ne  & -66    & -66   & -5   \\
$^{21}$Na  & -1.2   & -1.2  & -4   \\
$^{23}$Mg  & -4.5   & -4.7  & -3   \\
$^{25}$Al  & -1.6   & -2.4  & -3   \\
$^{27}$Si  & -4.6   & -6.5  & -2   \\
$^{29}$P   & -1.1   & -1.5  & -2   \\
$^{31}$S   & -7.3   & -7.8  & -1   \\
$^{33}$Cl  & 3.3   & 1.3   & -0.7 \\
$^{35}$Ar  & 2.6    & 1.8   & -0.4 \\
$^{37}$K   & -0.7   & -2.0  & -0.2 \\
$^{39}$Ca  & 3.7    & -0.0  & 0.1  \\
$^{41}$Sc  & 1.3    & -4.8  & 0.07 \\
$^{43}$Ti  & -5.1    & -9.0  & 0.4  \\
$^{45}$V   & -2.1    & -6.0  & 0.6 
\end{tabular}
\end{table}

For all isotopes up to $A = 23$, the shift due to the full recoil-correction, $\Delta A_{\beta, R}$, and the shift from the weak magnetism contribution, $\Delta A_{\beta, W\!M}$, alone (i.e. neglecting all form factor ratios except for $b/c_1$ in Eqs. \eqref{eq:H0} and \eqref{eq:H1}) are comparable, indicating that the recoil correction is indeed dominated by weak magnetism. For the heavier isotopes, especially the ones above $^{33}$Cl but also for $^{25}$Al and $^{27}$Si, the effect of the second order recoil correction, i.e. $\mathcal{O}(1/M^2)$, such as $g$, $a_2$ and $c_2$ becomes sizeable. 
Ref. \cite{WM_tbp} presented a non-zero, impulse approximation result for the $f$ form factor. If this second class form factor would be reinstated in Eq. \eqref{eq:H0} and Eq. \eqref{eq:H1} it would shift $\Delta A_{\beta, R}$ in the opposite direction as $g$.

\begin{figure}
    \centering
    \includegraphics[width=\columnwidth]{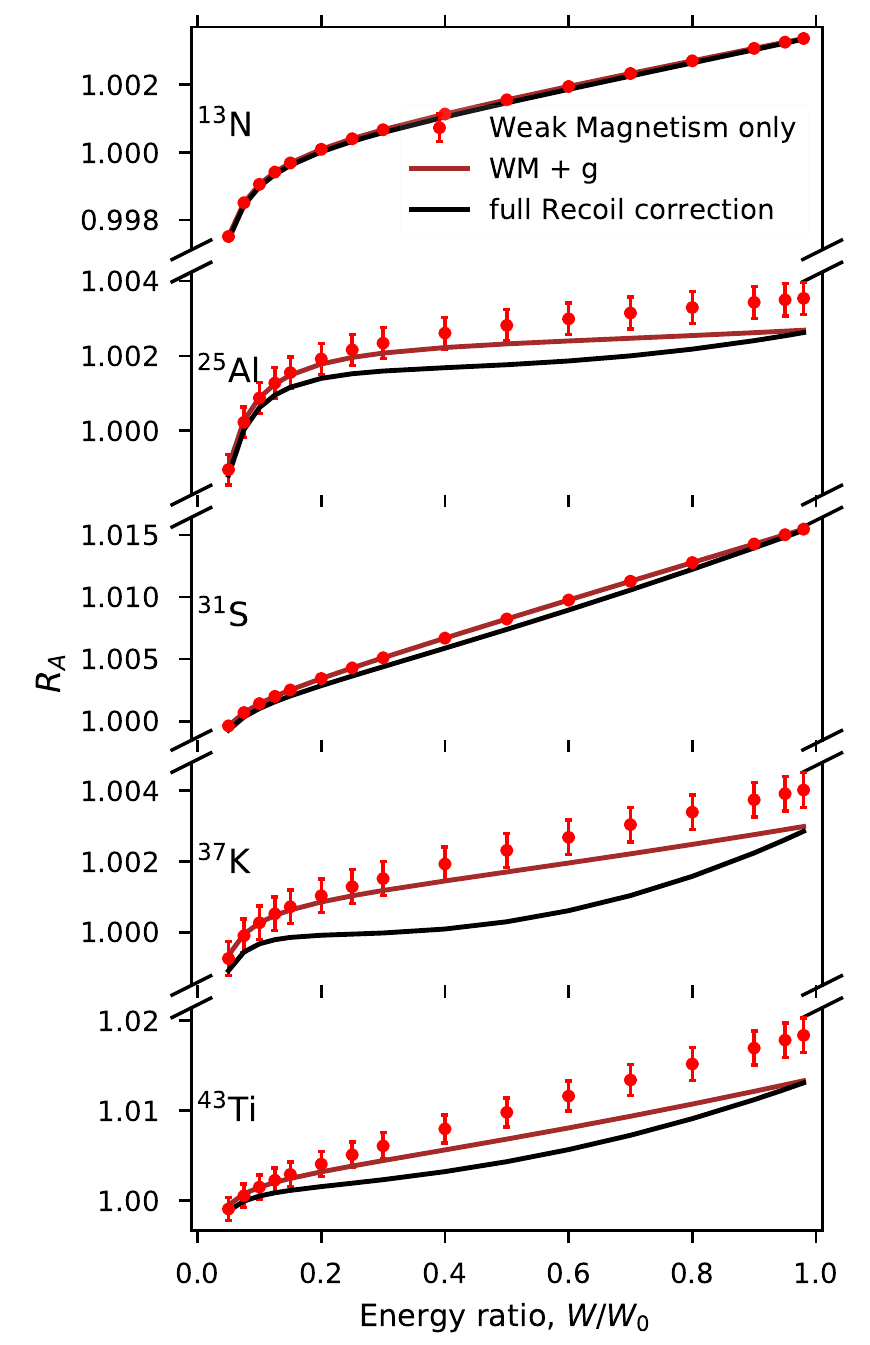}
    \caption{The correction ratio $R_A$ when including different parts of the recoil corrections for five different isotopes, i.e. $^{13}$N, $^{25}$Al, $^{31}$S, $^{37}$K, and $^{43}$Ti. Mind the different scales on the y-axis for the different isotopes.}
    \label{fig:overviewAbeta}
\end{figure}

Figure \ref{fig:overviewAbeta} shows the impact of the second order recoil corrections for five different isotopes, $^{13}$N, $^{25}$Al, $^{31}$S, $^{37}$K, and $^{43}$Ti. For every isotope, it compares the energy dependence of $R_A$ when including either only the weak magnetism form factor, weak magnetism and the $g$ form factor, or the full recoil correction. The importance of the $g$ form factor depends on the spin of the isotope. Being zero for $J = 1/2$, i.e. for $^{13}$N and $^{31}$S, whereas it leads to an important contribution for e.g. $^{43}$Ti. For $^{37}$K, with with recoil-order corrections on the permille level only, the influence of $a_2$ and $c_2$ is seen to dominate.

\subsection{Radiative correction}
\begin{figure}
	\centering
	\includegraphics[width=\columnwidth]{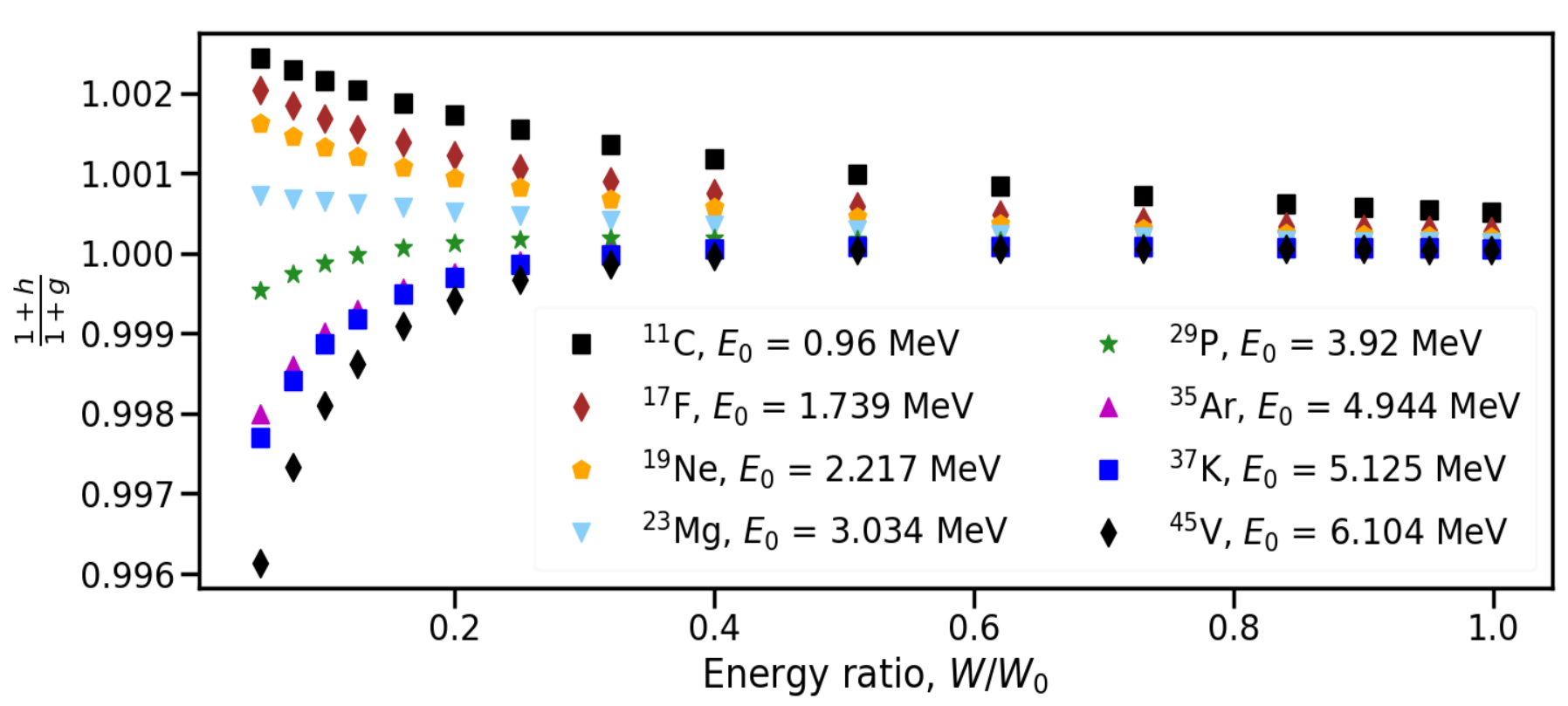}
	\caption{Radiative correction to the beta-asymmetry parameter as described in Eq. \eqref{eq:A_RC} for some selected mirror nuclei. Energies are given as the ratio between the decay energy, $W$, and the endpoint energy, $W_0$. The influence is seen to be small throughout the energy range and stays well below $10^{-3}$ except for the lowest energies. Up to endpoint energies $W_0=3.5$ MeV, the size of the correction decreases but for higher endpoint energies the sign of the correction flips and its influence increases again.}
	\label{fig:A_RC}
\end{figure}

The radiative correction on the beta-asymmetry parameter $A_{\beta}$ is calculated with the analytical expressions presented in Section \ref{sec:intro_rc}. The beta-asymmetry parameter, with radiative corrections included, is easily obtained from Eq. \eqref{eq:N_asymmetry} as
\begin{equation} \label{eq:A_RC}
    A_{\beta, RC} = A_{\beta, 0} \times \frac{1 + \frac{\alpha}{2\pi} h(\beta)}{1 + \frac{\alpha}{2\pi} g(\beta)},
\end{equation}
with $A_{\beta, 0}$ given in Eq. \eqref{eq:A_0}. 
This equation allows one to easily calculate the importance of radiative corrections on the beta asymmetry and highlights the fact that it is the ratio of the correction functions that has an impact. \\
For some selected nuclei the size of the correction as calculated using Eq. \eqref{eq:A_RC} is shown in Fig. \ref{fig:A_RC}. In all previous studies listed in Table \ref{tab:meas}, except for the most recent one \cite{new19Ne}, the radiative correction has been omitted in the analysis. The effect is shown to not exceed $2 \times 10^{-3}$ and even stays well below $10^{-3}$ above the first 10 \% of the beta decay energy spectrum. In conclusion, neglecting the radiative corrections was justified, with the only possible exception of $^{19}$Ne. Remember that for $^{19}$Ne the leading order expression is suppressed and all corrections should therefore be considered carefully.

\begin{figure}
	\centering
	\includegraphics[width=\columnwidth]{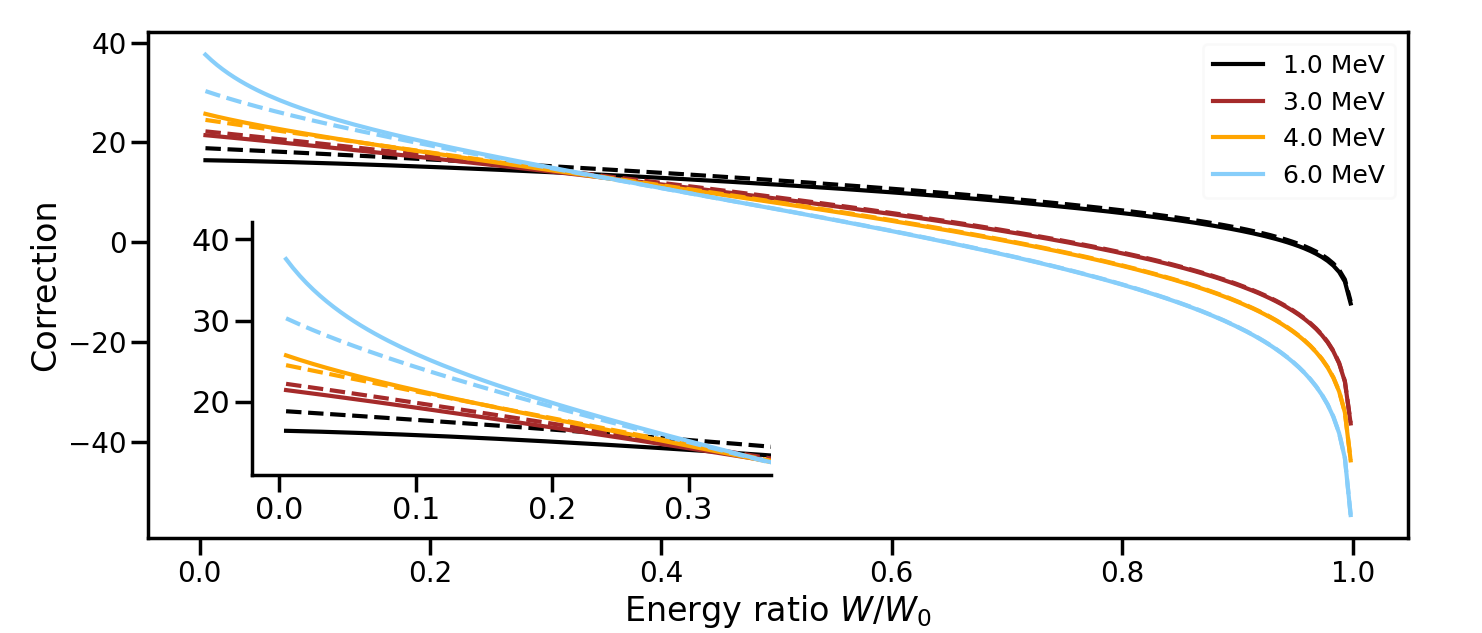}
	\caption{Size of the correction function $g(W_0,W)$, Eq. \eqref{eq:gSirlin}, (solid) and $h(W_0, W)$, Eq. \eqref{eq:h}, (dashed) for different endpoint energies. For all endpoints the difference between both is most pronounced at lower energy ratios, $W/W_0$, but their relative size changes with respect to each other. This behaviour is the cause of the sign flip in the radiative correction as illustrated in Fig. \ref{fig:A_RC}.}
	\label{fig:g_vs_h}
\end{figure}

Both the size and the sign of the radiative correction depend on the endpoint energy of the considered decay. For low endpoint energies, $h(W_{0}, W)$ is larger than $g(W_{0}, W)$ and the radiative correction increases the size of the correlation coefficient. The difference between both functions decreases for higher endpoint energies and reaches a minimum around $3.5$ MeV, with minimal impact of the radiative corrections. For higher endpoint energies, the difference increases again, but this time $g(W_0, W)$ is larger, accordingly, reducing the beta-asymmetry parameter. This trend can clearly be observed in Fig. \ref{fig:g_vs_h}. In Fig. \ref{fig:A_RC}, the resulting effect on the radiative correction is shown for several mirror beta decays with different endpoint energies. Note that the trend resembles a $1/W$ dependence, implying a per-mille level Fierz interference term should one neglect the radiative correction.\\
Using an idealised experimental analysis, identical to the previous section, the induced systematic error is deduced. The systematic error when neglecting the radiative correction, $\Delta A_{\beta,RC}$, compared to $A_{\beta,0}$ is shown in Table \ref{tab:offset}. For the mirror nuclei under consideration, the influence is clearly diminishing with atomic number, and thus with endpoint energy.  \\
For most mirror nuclei, the effect of the radiative corrections on the beta-asymmetry parameter is negligible at the current level of precision and much smaller than in, for example, the $\beta$ spectrum shape. This conclusion could have been expected from Eq. \eqref{eq:A_RC}. Indeed, the size of the correction is diminished due to the appearance of a ratio between two similar expressions. 

\section{Beta-neutrino correlation} \label{sec:betaneut}
The angular correlation between the beta particle and the neutrino cannot be measured directly due to the elusive nature of the neutrino. Experimentally, the correlation is either inferred from a measurement of the daughter nucleus recoil energy \cite{PhysRev116, Lienard2015} or by observing beta-delayed emission of secondary particles \cite{21Na, Egorov1997, Adelberger1999}. Neglecting radiative beta decay, no difference exists between the theoretical and experimental beta-neutrino correlation, as both are three-body decays. However, including radiative beta decay, $p \rightarrow n e^-\nu \gamma$, changes the process from a three- to a four-body decay process, and as such the phase space is changed. 

How the specific experimental situation impacts the analysis of correlation coefficient measurements is discussed in more detail in Ref. \cite{Hayen_tbp, Gluck_tbp}. Here, we describe the theoretical framework for the effect of the recoil and radiative corrections, give examples for idealised experiments and provide previously unpublished semi-analytical values to calculate the radiative correction for realistic experimental conditions. These semi-analytical values significantly expand the information available to date.

\subsection{Recoil effect}
In the Holstein formalism the beta-neutrino correlation $a_{\beta \nu}$ including recoil terms, is written as:
\begin{equation}
	a_{\beta \nu, R} = \frac{F_2 + \Delta F_2}{H_0 + \Delta F_1}
\end{equation}
with $H_0$ (Eq. \eqref{eq:H0}), $F_2$, $\Delta F_1$ (Eq. \eqref{eq:deltaF1}) and $\Delta F_2$ again from the appendices of Ref. \cite{Holstein1974}.
For $F_2$, one has,
\begin{equation} \label{eq:F2}
\begin{aligned}
  F_2 (W) = & 1 - 2\frac{a_2}{a_1} - \frac{2}{3}\frac{c_2}{c_1} (1 + 8WW_0 - 8W^2) \\
  &- \frac{1}{3} \left(\frac{c_1}{a_1}\right)^2 \bigg[ 1 - \frac{2W_0}{M} \left(1 \pm \frac{b}{c_1} \right)\\
  & + \frac{4W}{M} \left(3\pm \frac{b}{c_1} \right) \bigg],
\end{aligned}
\end{equation}
and for the electromagnetic modifications,
\begin{equation} \label{eq:deltaF2}
\begin{aligned}
  \Delta F_2(W) = \mp \frac{8 \alpha Z}{ 3 \pi} & \bigg[ \bigg( 8 W + W_0 \bigg) \\
  & - \rho^{2} ( 8W - W_0 ) \bigg]X.
\end{aligned}
\end{equation}
An important difference with the expression for the beta-asymmetry, discussed in Section \ref{sec:asymmetry}, is the absence of the higher order form factor $g$, one of the $\mathcal{O}(1/M^2)$ form factors which were found to play a significant role in the analysis of $A_{\beta}$ for $A>25$. 

\begin{figure}
	\centering
	\includegraphics[width=\columnwidth]{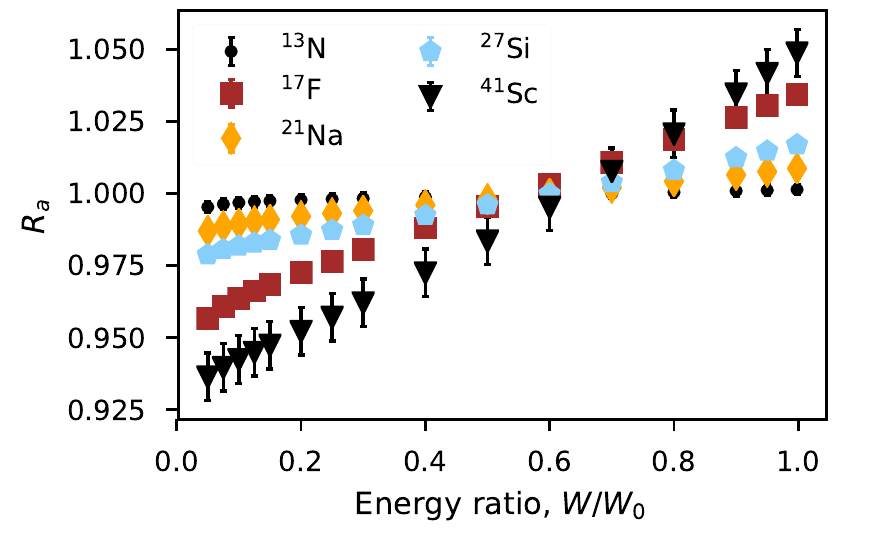}
	\caption{The influence of the recoil correction terms for mirror nuclei with different endpoint energies. The size of the correction shows  no trend with respect to the endpoint energy. Error bars include the uncertainty from Ref. \cite{WM_tbp} on the end point energy, the weak magnetism form factor $b_{W\!M}$, and the Gamow-Teller to Fermi strength ratio $\rho$. The latter are the dominant contribution.}
	\label{fig:a_recoil}
\end{figure}

\begin{figure}
    \centering
    \includegraphics[width=\columnwidth]{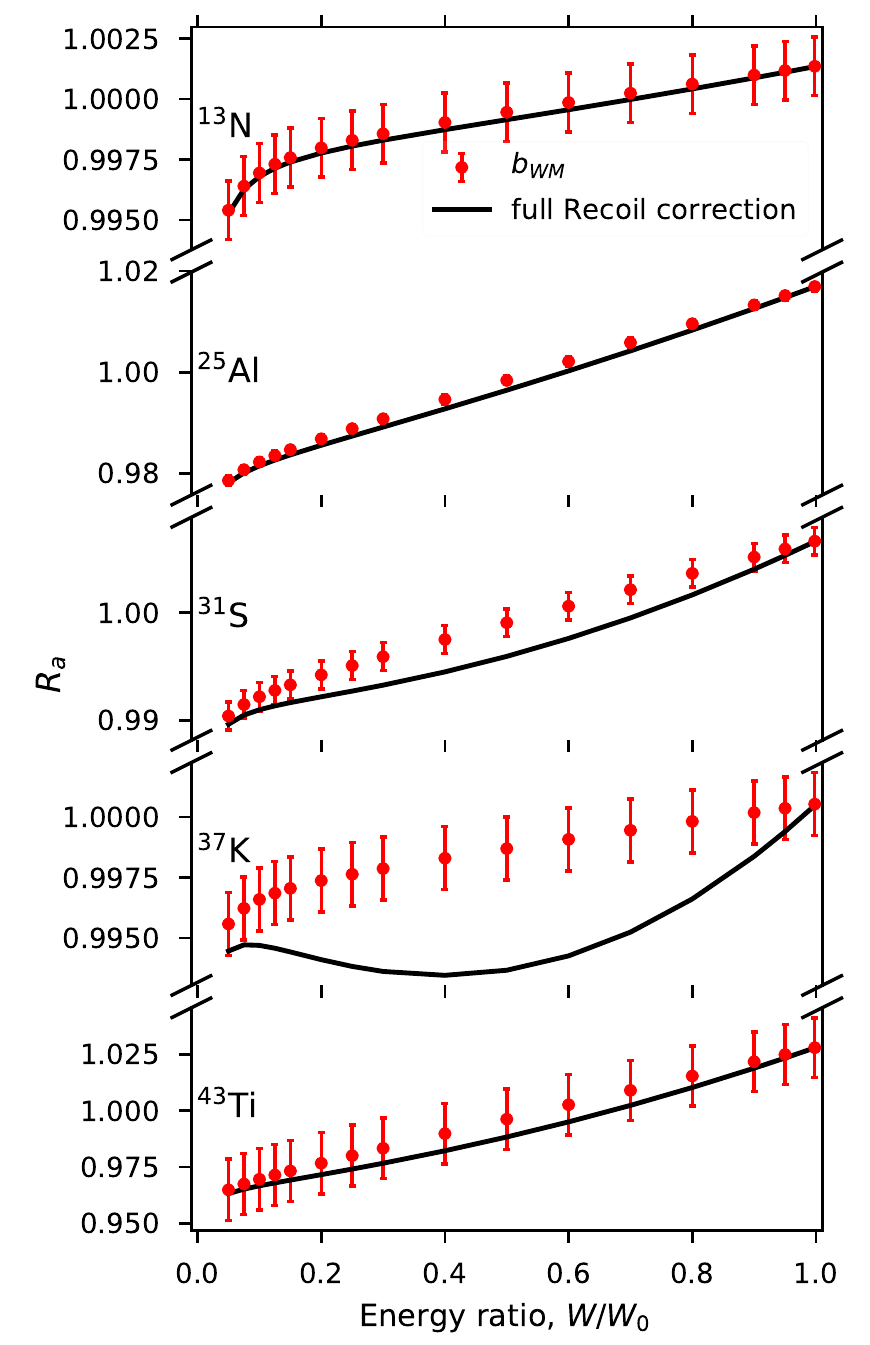}
    \caption{The correction ratio $R_a$ when including either only the weak magnetism form factor or the full recoil corrections for five different isotopes, $^{13}$N, $^{25}$Al, $^{31}$S, $^{37}$K, and $^{43}$Ti. Mind the different scales on the y-axis for the different isotopes.}
    \label{fig:overview_abnu}
\end{figure}

The impact of the recoil correction throughout the energy spectrum is defined analogous to Eq. \eqref{eq:R_A} as:
\begin{equation}
R_a = \frac{a_{\beta \nu, R}}{a_{\beta \nu, 0}},
\end{equation}
and is shown in Fig. \ref{fig:a_recoil}.
Similar to the recoil corrections on the $\beta$-asymmetry, no trend is observed in the size of the correction terms with increasing endpoint energy. The correction shows a dominantly linear behaviour with more pronounced deviations from this trend at lower energies, i.e. $W/W_0<0.2$, for low endpoint transitions. 
The impact of the second order recoil corrections in again visualised for five different isotopes, $^{13}$N, $^{25}$Al, $^{31}$S, $^{37}$K, and $^{43}$Ti, in Fig. \ref{fig:overview_abnu}. For every isotope, it compares $R_a$ including only the weak magnetism form factor with respect to the full recoil correction.
For the beta-neutrino correlation coefficient the weak magnetism form factor dominates the recoil-order corrections except for $^{33}$Cl, $^{35}$Ar, $^{37}$K, and $^{39}$Ca. All of the latter are found to have weak magnetism form factors $b/Ac<2.2$ \cite{WM_tbp}.

For isotopes in the tails of Fig. \ref{fig:abnu_sensi}, with small values for $a_{\beta \nu, 0}$, the relative importance of the recoil-order corrections increases, most notably for $^{19}$Ne. The correction ratio $R_a$ for $^{19}$Ne was shown in Fig. \ref{fig:Ne_corr} already. Its impact is seen to be larger than for the beta-asymmetry parameter but behaves similarly. Also for several other isotopes towards the tail of Fig. \ref{fig:abnu_sensi}, i.e. $^{17}$F, $^{41}$Sc, $^{25}$Al and $^{21}$Na, the relative importance of the recoil correction $R_a$ is found to be larger than $R_A$.

\subsection{Radiative correction} 
As Gl\"uck and collaborators \cite{Toth1986, Gluck1990, Gluck1997, Gluck_tbp} have shown, former calculations of the radiative correction to correlations coefficients involving the (anti-)neutrino momentum, such as the beta-neutrino correlation $a_{\beta\nu}$ or the neutrino asymmetry $B_{\nu}$ \cite{Shann1971, YM1973, YM1976, GM1978}, are incorrect to use in an experimental analysis. Measurements determine the correlation between the beta particle and the recoiling nucleus to avoid the challenging detection of the neutrino. However, both are no longer interchangeable when an external photon is emitted.
When the shift to a four body-decay is overlooked, this omission influences the experimental results unless the neutrino or the additional photon is detected. 

The main shortcoming of previous analytical calculations \cite{YM1976,GM1978,Ando2004,Ivanov2013} resides in the bremsstrahlung photon integration, which is performed with fixed neutrino direction. In that case, due to momentum conservation, the recoil particle momentum follows the photon momentum, and as a consequence the information on the recoil particle is lost during the integration. The alternative, and correct, procedure is to perform the photon integration by fixing the recoil particle momentum. In this case, the neutrino momentum follows the photon momentum, and only the information on the neutrino momentum is lost. Unfortunately, the integrations in the latter case are more difficult. This procedure, first introduced by Gl\"uck and collaborators, can be performed using an additional energy splitting with respect to the one already incorporated by Sirlin. The energy of the photon is split between soft and hard photons with a cut-off energy defined by $\omega$. When the virtual photon exchange and the inner bremsstrahlung diagrams are calculated separately an infrared divergent result is obtained. The proposed splitting allows the soft photon calculation to confine the divergence while only having a very small kinematic impact. For the hard photons, however, the deviation from a three-body decay becomes important, and a separate calculation is implemented. The effect can be estimated using semi-analytical integrals \cite{Gluck1993, SANDI} or calculated using a Monte Carlo routine \cite{Gluck1997, GENDER} for a specific experimental set-up.

The implementation and results for the specific cases of the neutron \cite{Gluck1993} and the recoil spectra of $^6$He and $^{32}$Ar \cite{Gluck1998} have been published before, resulting in a radiative correction shift of $\approx3\times10^{-3}$ on $a_{\beta\nu}$ for $^{6}$He. In the present study, we expand this to the recoil spectrum of the mirror $\beta$ decays up to $A=45$. The performance of the new calculations has been tested by comparing to the already published results. A more in-depth explanation on the method is given in Refs. \cite{Gluck1992, Gluck1993, Gluck_tbp, SANDI}. The results for the recoil spectra are given in Table \ref{tab:a_semianalytical}. 
Using the relative radiative corrections $r(T)$, listed there in percentages, the corrected recoil spectrum $R(T)$ including the order-$\alpha$ radiative correction
can be obtained as 
\begin{equation} \label{eq:inclusion_radcorr}
 \begin{aligned}
  R(T) & = R_{0C}(T)+R_{\gamma C}(T) \\ 
  & = R_{0C}(T)\cdot \left(1+\frac{r(T)+r_{\rm tot}}{100}\right),
 \end{aligned}
\end{equation}
with $T$ the recoil kinetic energy, $R_{0C}(T)$ the recoil spectrum calculated to zeroth order (i.e. without recoil-order or radiative corrections) but with Coulomb correction (Fermi function), and $r_{\rm tot}$ denoting the relative radiative correction to the total decay rate defined as
\begin{equation} \label{eq:rtot}
r_{\rm tot}=100 \frac{\int dT R_{\gamma C}(T)}{\int dT R_{0C}(T)}.
\end{equation}
The zeroth order recoil spectrum can be calculated by integrating the zeroth-order decay rate distribution (also called Dalitz distribution)
\begin{equation}
 \begin{aligned}
 w_{0C}(W,T)= & \frac{M(1+\rho^2)}{4\pi^3}F_0(W)
 \big[ W_\nu W(1+a_{\beta\nu}) \\
 & + a_{\beta\nu}M(T-T_0)\big] 
 \end{aligned}
\end{equation}
with respect to the beta energy $W$ (see Eqs. A.7-A.13 in Ref. \cite{Gluck1998}); $T_0$ denotes the maximum recoil kinetic energy.
The radiative correction recoil spectrum $R_{\gamma C}(T)$ is computed by integrating the radiative correction decay rate $w_{\gamma C}(W,T)$ with respect to $W$. Here the subindex $C$ means that we multiply both the order-$\alpha$ virtual and bremsstrahlung correction functions by the Fermi function $F_0(W)$. Doing so the order-$Z\alpha^2$ radiative correction is also partly taken into account \cite{Gluck1998}.
The semianalytical bremsstrahlung photon integration method to get $w_{\gamma C}(W,T)$ is explained in the appendix of Ref. \cite{Gluck1993}. A more detailed description of this calculation will be given in an upcoming publication \cite{SANDI}.

\begin{figure}
	\centering
	\includegraphics[scale=.23]{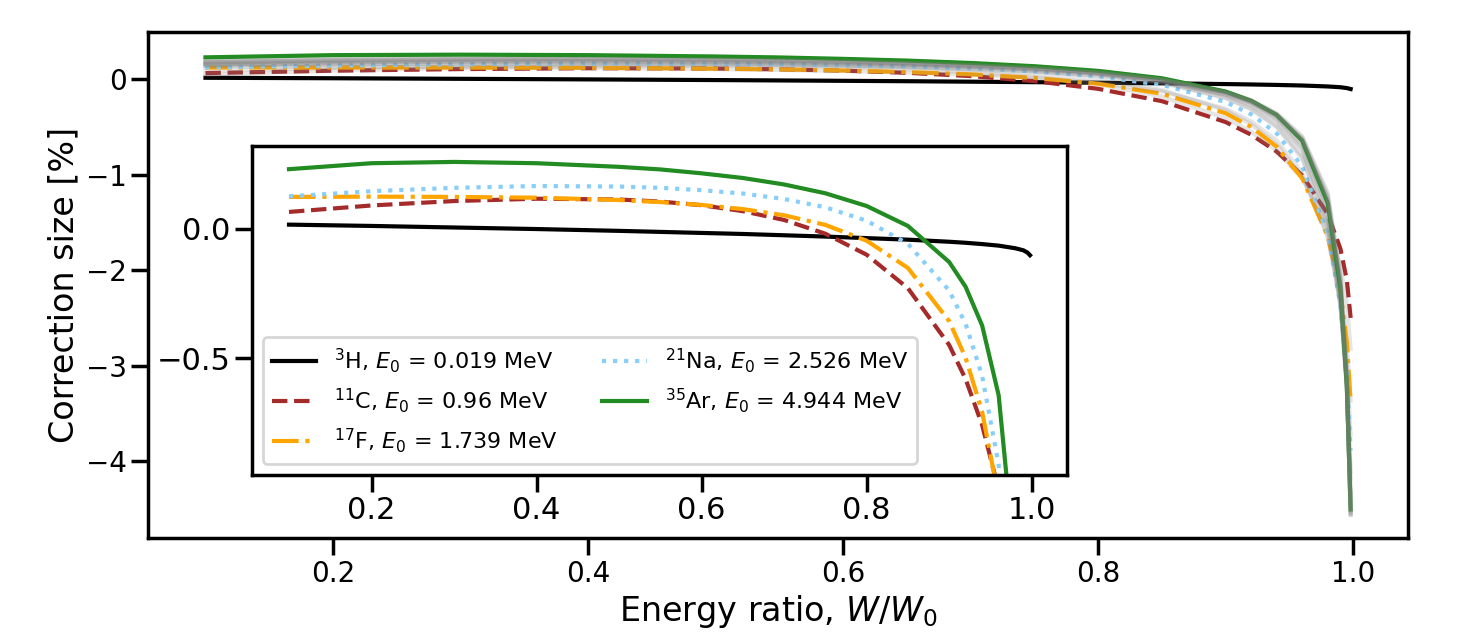}
	\caption{The radiative correction, $r(T)$, to the recoil spectrum in the beta-neutrino correlation $a_{\beta \nu, RC}$ as calculated using semi-analytical integrals \cite{Gluck1993, SANDI} for several mirror decays. The correction is most important for high energy ratios $T/T_0$. The correction is very similar for all decays as is also visible in the inset, with the exception of $^{3}$H due to its low endpoint energy. Numerical values are tabulated in Table \ref{tab:a_semianalytical}.}
	\label{fig:a_semianaly}
\end{figure}

For several already performed and planned experiments, listed in Table \ref{tab:meas}, the correction is also plotted in Fig. \ref{fig:a_semianaly}. The impact of the radiative corrections is approximately constant at  $0.1\%$ up to an energy ratio of about $T/T_0=0.7$. Close to the endpoint the impact increases significantly, which is opposite to results shown in Fig. \ref{fig:A_RC} for the beta asymmetry. However, it should be noted that this remarkable increase bears a close resemblance to the unphysical logarithmic divergence known to exist in Eq. \eqref{eq:gSirlin} \cite{Repko1983, Gardner2004, Gluck_tbp}. Proposed solutions are to either sum the contribution of soft real photons to all orders \cite{Repko1983}, or take into account the non-zero energy resolution of the electron detector \cite{Gardner2004}. 
Another prominent characteristic in Fig. \ref{fig:a_semianaly} is the similarity of the radiative correction between different decays. The dependence of the recoil spectrum radiative correction on the endpoint energy of the transition is indeed small, in contrast to the strong dependence of the electron spectrum on $W_0$ (See Figs \ref{fig:A_RC} and \ref{fig:g_vs_h}). This result is connected with the Kinoshita-Lee-Nauenberg theorem \cite{Kinoshita, Lee_Nauenberg}. Due to the collinear peaks of the Feynman amplitudes when the photon is collinear with the electron the radiative corrections to observables with non-integrated electron energy have strong electron energy, $W$, and endpoint energy, $W_0$, dependence.
On the other hand, the radiative corrections to observables with integrated electron energy (like the total decay rate, the recoil spectrum or the neutrino energy spectrum) do not have this strong dependences and are finite in the $m_e\rightarrow0$ limit \cite{Sirlin2013}. 

In conclusion, the older calculations assumed (implicitly) that the neutrino rather than the recoiling nucleus would be detected, whereas the here-presented results assume it is never detected.
Note that the radiative correction results depend very much on the experimental details. For example, if the recoil particle energy spectrum is measured in coincidence with the beta particle (Dalitz distribution), one has to use completely different radiative corrections \cite{Gluck1993, Gluck_tbp} than those presented in the present paper.
Therefore, the preferred method to take radiative corrections into account for a given experimental set-up is by implementing a Monte Carlo routine \cite{Gluck1997, GENDER}. 

\section{Conclusion}

The determination of the $\beta$ decay correlation coefficients can provide bounds on the existence of exotic currents, which are possible extensions of the present weak interaction description included in the SM. In this context, the mirror transitions provide crucial input towards the possible existence of right-handed neutrinos \cite{Falkowski2021}. We have focused on the mirror beta decays up to $A=45$ and discussed the experimentally most relevant correlations, the beta-asymmetry and the beta-neutrino angular correlation. More specifically, the leading-order SM values have been updated using the most recent results for $\rho$ \cite{WM_tbp} which significantly changed the value of $A_{\beta}$ for $^{3}$H. Furthermore, the importance of the inclusion of two higher-order correction terms, i.e. the recoil-order and radiative correction, has been discussed. 

The recoil-order correction is the largest of the two here-discussed corrections for both the beta asymmetry and the beta-neutrino correlation. Its impact has been evaluated starting from a recent evaluation of the form factors for the mirror beta-decays \cite{WM_tbp}. Using the there summarised experimental weak magnetism form factor and calculations within the impulse approximation it has been concluded that the form factors of $\mathcal{O}(1/M^2)$ play a significant role in the beta-asymmetry parameter for $^{25}$Al, $^{27}$Si, and the mirror decays with $A>31$. For the beta-neutrino correlation the weak magnetism form factor dominates the recoil-order correction except when its value is small, i.e. for $^{33}$Cl,  $^{35}$Ar,  $^{37}$K, and  $^{39}$Ca.

The evaluation of the radiative corrections differs between both observables. For the beta-asymmetry, the evaluation can rely on analytical results, while for the beta-neutrino angular correlation this is not possible. Measurements do not determine the correlation between the beta particle and the neutrino directly but rather reconstruct it from a detection of the recoiling nucleus momentum. When this reconstruction overlooks the shift to a four body-decay induced by radiative beta decay the obtained result is not correct. To take complicated experimental details into account, Monte-Carlo methods are essential for the radiative correction calculations. Nevertheless, the semi-analytical results presented here allow for an estimate of their effect. 
 
Our broader study examining, and comparing, the sensitivity to new physics as well as the size of the corrections for a range of different isotopes provides valuable information for planning and comparing future experimental efforts. 
Depending on the experimental sensitivity the here-discussed corrections will have to be included. For an experimental sensitivity at the percent level, the recoil corrections should be included. Higher order form factors become important when the experimental sensitivity reaches $10^{-3}$ especially for isotopes with $A>26$. Only at permille precision, the radiative corrections come into play.
For nuclei with similar sensitivity (see Table \ref{tab:sensitivity}) the availability of the weak magnetism form factor and the number of additional form factors, together with experimental conditions, will help choosing the most suitable isotope for an experiment.

\begin{turnpage}
\begin{table*}
\caption{Size of the relative radiative correction to the total decay rate, $r_{tot}$, and the radiative correction, $r(T)$, presented in Eqs. \eqref{eq:rtot} and \eqref{eq:inclusion_radcorr}, respectively. These values are calculated using semi-analytical integrals \cite{Gluck1993, SANDI} for the beta-neutrino angular correlation $a_{\beta \nu}$ in the mirror decays for different values of the energy ratio, $y=T/T_0$. }
\label{tab:a_semianalytical}
\begin{tabular}{c | r |rrrrrrrrrrrrrrrrrrrr}
\hline \hline
Parent & \multicolumn{1}{c}{$r_{tot}$} & \multicolumn{20}{c}{$r(T)$} \\ 

 & & \multicolumn{1}{c}{0.1} & \multicolumn{1}{c}{0.2} &
  \multicolumn{1}{c}{0.3} & \multicolumn{1}{c}{0.4} &
  \multicolumn{1}{c}{0.5} & \multicolumn{1}{c}{0.55} &
  \multicolumn{1}{c}{0.6} & \multicolumn{1}{c}{0.65} &
  \multicolumn{1}{c}{0.7} & \multicolumn{1}{c}{0.75} &
  \multicolumn{1}{c}{0.8} & \multicolumn{1}{c}{0.85} &
  \multicolumn{1}{c}{0.9} & \multicolumn{1}{c}{0.92} &
  \multicolumn{1}{c}{0.94} & \multicolumn{1}{c}{0.96} &
  \multicolumn{1}{c}{0.98} & \multicolumn{1}{c}{0.99} &
  \multicolumn{1}{c}{0.995} & \multicolumn{1}{c}{0.998} \\ \hline
$^{3}$H & 1.816 & 0.013 & 0.008 & 0.002 & -0.004 & -0.011 & -0.015 & -0.019 & -0.024 & -0.028 & -0.033 & -0.039 & -0.045 & -0.053 & -0.058 & -0.062 & -0.068 & -0.078 & -0.087 & -0.095 & -0.105 \\
$^{11}$C & 1.450 & 0.068 & 0.091 & 0.106 & 0.113 & 0.108 & 0.102 & 0.089 & 0.069 & 0.036 & -0.015 & -0.095 & -0.223 & -0.441 & -0.573 & -0.749 & -0.999 & -1.404 & -1.768 & -2.095 & -2.491 \\
$^{13}$N & 1.396 & 0.051 & 0.092 & 0.119 & 0.134 & 0.135 & 0.129 & 0.120 & 0.103 & 0.074 & 0.028 & -0.047 & -0.173 & -0.399 & -0.542 & -0.738 & -1.027 & -1.510 & -1.948 & -2.341 & -2.812 \\
$^{15}$O & 1.298 & 0.090 & 0.120 & 0.139 & 0.150 & 0.150 & 0.145 & 0.135 & 0.120 & 0.098 & 0.062 & 0.001 & -0.107 & -0.318 & -0.463 & -0.675 & -1.011 & -1.622 & -2.207 & -2.735 & -3.357 \\
$^{17}$F  & 1.295 & 0.125 & 0.125 & 0.123 & 0.118 & 0.107 & 0.098 & 0.086 & 0.070 & 0.048 & 0.014 & -0.043 & -0.145 & -0.348 & -0.488 & -0.696 & -1.027 & -1.633 & -2.217 & -2.745 & -3.368 \\
$^{19}$Ne & 1.225 & 0.135 & 0.131 & 0.126 & 0.118 & 0.104 & 0.095 & 0.082 & 0.065 & 0.043 & 0.013 & -0.036 & -0.123 & -0.302 & -0.431 & -0.632 & -0.970 & -1.641 & -2.329 & -2.965 & -3.712 \\
$^{21}$Na & 1.185 & 0.127 & 0.146 & 0.158 & 0.163 & 0.160 & 0.154 & 0.145 & 0.131 & 0.109 & 0.078 & 0.030 & -0.056 & -0.231 & -0.359 & -0.557 & -0.900 & -1.604 & -2.351 & -3.053 & -3.876 \\
$^{23}$Mg & 1.129 & 0.140 & 0.167 & 0.182 & 0.188 & 0.184 & 0.178 & 0.168 & 0.153 & 0.131 & 0.099 & 0.051 & -0.029 & -0.189 & -0.306 & -0.492 & -0.822 & -1.542 & -2.359 & -3.155 & -4.096 \\
$^{25}$Al & 1.106 & 0.144 & 0.157 & 0.164 & 0.166 & 0.161 & 0.155 & 0.146 & 0.132 & 0.111 & 0.080 & 0.035 & -0.040 & -0.191 & -0.302 & -0.479 & -0.799 & -1.516 & -2.352 & -3.184 & -4.173 \\
$^{27}$Si & 1.058 & 0.155 & 0.172 & 0.181 & 0.184 & 0.178 & 0.172 & 0.162 & 0.147 & 0.126 & 0.095 & 0.049 & -0.024 & -0.166 & -0.269 & -0.434 & -0.735 & -1.439 & -2.312 & -3.219 & -4.319 \\
$^{29}$P & 1.047 & 0.165 & 0.189 & 0.201 & 0.204 & 0.198 & 0.190 & 0.179 & 0.164 & 0.141 & 0.109 & 0.062 & -0.013 & -0.155 & -0.257 & -0.420 & -0.717 & -1.418 & -2.297 & -3.220 & -4.346 \\
$^{31}$S & 1.011 & 0.176 & 0.198 & 0.209 & 0.211 & 0.203 & 0.195 & 0.184 & 0.168 & 0.145 & 0.113 & 0.066 & -0.008 & -0.145 & -0.242 & -0.396 & -0.677 & -1.357 & -2.247 & -3.218 & -4.428 \\
$^{33}$Cl & 0.998 & 0.210 & 0.237 & 0.245 & 0.242 & 0.230 & 0.221 & 0.207 & 0.188 & 0.164 & 0.130 & 0.080 & 0.004 & -0.134 & -0.230 & -0.382 & -0.659 & -1.331 & -2.223 & -3.211 & -4.454 \\
$^{35}$Ar & 0.972 & 0.230 & 0.252 & 0.257 & 0.252 & 0.238 & 0.227 & 0.212 & 0.193 & 0.168 & 0.134 & 0.084 & 0.008 & -0.128 & -0.221 & -0.367 & -0.634 & -1.285 & -2.176 & -3.191 & -4.497 \\
$^{37}$K & 0.960 & 0.183 & 0.202 & 0.211 & 0.212 & 0.204 & 0.196 & 0.184 & 0.168 & 0.145 & 0.114 & 0.067 & -0.006 & -0.137 & -0.228 & -0.370 & -0.629 & -1.269 & -2.155 & -3.180 & -4.514 \\
$^{39}$Ca & 0.937 & 0.179 & 0.195 & 0.203 & 0.203 & 0.196 & 0.188 & 0.177 & 0.162 & 0.140 & 0.108 & 0.062 & -0.010 & -0.138 & -0.226 & -0.363 & -0.613 & -1.231 & -2.108 & -3.151 & -4.539 \\
$^{41}$Sc & 0.939 & 0.158 & 0.162 & 0.164 & 0.162 & 0.154 & 0.147 & 0.137 & 0.123 & 0.103 & 0.074 & 0.031 & -0.037 & -0.160 & -0.245 & -0.380 & -0.626 & -1.242 & -2.116 & -3.157 & -4.539 \\
$^{43}$Ti & 0.919 & 0.171 & 0.181 & 0.187 & 0.187 & 0.180 & 0.173 & 0.163 & 0.148 & 0.127 & 0.097 & 0.052 & -0.018 & -0.144 & -0.229 & -0.362 & -0.603 & -1.203 & -2.068 & -3.122 & -4.553 \\
$^{45}$V & 0.905 & 0.187 & 0.203 & 0.210 & 0.211 & 0.202 & 0.194 & 0.183 & 0.167 & 0.145 & 0.113 & 0.067 & -0.006 & -0.133 & -0.219 & -0.351 & -0.589 & -1.179 & -2.036 & -3.097 & -4.559 \\

\end{tabular}
\end{table*}
\end{turnpage}

\newpage

\bibliographystyle{unsrt}
\bibliography{References.bib}

\end{document}